\documentclass[showpacs,twocolumn,amsmath,superscriptaddress,prl]{revtex4}
\usepackage{epsfig}
\usepackage{epstopdf} 
\usepackage{graphicx}
\begin{document} 
\title{Statistical similarity between the compression of a porous
material and earthquakes} 
\author{Jordi Bar\`o}
\email{jordibaro@ecm.ub.es} 
\affiliation{Departament d'Estructura i Constituents de la Mat\`eria,
Facultat de F\'{\i}sica, Universitat de Barcelona, Mart\'{\i} i
Franqu\`{e}s 1, E-08028 Barcelona, Catalonia.}
\author{\'Alvaro Corral}
\affiliation{Centre de Recerca Matem\`atica,
Edifici C, Campus Bellaterra,
E-08193 Bellaterra, Spain.
} 
\author{Xavier Illa} 
\affiliation{Departament d'Estructura i Constituents de la Mat\`eria,
Facultat de F\'{\i}sica, Universitat de Barcelona, Mart\'{\i} i
Franqu\`{e}s 1, E-08028 Barcelona, Catalonia.} 
\author{Antoni Planes} 
\affiliation{Departament d'Estructura i Constituents de la Mat\`eria,
Facultat de F\'{\i}sica, Universitat de Barcelona, Mart\'{\i} i
Franqu\`{e}s 1, E-08028 Barcelona, Catalonia.} 
 \author{Ekhard K. H. Salje}
\affiliation{Department of Earth Sciences, University of Cambridge,
Downing Street, Cambridge CB2 3EQ, UK.} 
\author{Wilfried Schranz} 
\affiliation{Faculty of Physics, University of Viena,
Boltzmanngasse 5, Vienna A-1090, Austria.} 
\author{Daniel E. Soto-Parra} 
\affiliation{Divisi\'{o}n de Materiales Avanzados, IPICYT,
Camino a la Presa San Jos\'{e} 2055, San Lu\'{\i}s Potos\'{\i},
M\'exico}
 \author{Eduard Vives} 
\affiliation{Departament d'Estructura i Constituents de la Mat\`eria,
Facultat de F\'{\i}sica,
Universitat de Barcelona, Mart\'{\i} i Franqu\`{e}s 1, E-08028
Barcelona, Catalonia.} 
\begin{abstract} 
It has been long stated that there are profound analogies between
fracture experiments and earthquakes; however, few works attempt a
complete characterization of the parallelisms between these so separate
phenomena. We study the Acoustic Emission events produced during the
compression of Vycor (SiO$_2$). The Gutenberg-Richter law, the modified
Omori's law, and the law of aftershock productivity hold
for a minimum of 5 decades, are independent of the compression rate, and
keep stationary for all the duration of the experiments. The
waiting-time distribution fulfills a unified scaling law with a
power-law exponent close to 2.45 for long times, which is explained in
terms of the temporal variations of the activity rate.
\end{abstract} 


\maketitle

Mechanical failure of materials is a complex phenomenon underlying many
accidents and natural disasters ranging from the fracture of small
devices under fatigue to earthquakes. Despite the vast separation of
spatial, temporal, energy, and strain-rate scales
\cite{Ben_Zion_review,Main}, and the differences in geometry, boundary
conditions, loading, structure of the medium, and interactions, it has
been proposed that laboratory experiments on brittle fracture in
heterogeneous materials can be a model for earthquake occurrence
\cite{Mogi62,Scholz68,Bonamy}. As the main stresses on Earth's crust are
compressive \cite{Main}, experiments of materials loaded under
compression seem the most suitable to draw analogies with seismicity.
But due to the fact that compression stabilizes crack propagation,
traditional assumptions applied to samples loaded under tension are not
valid in compression, making the compression problem much more
challenging conceptually \cite{Girard}.

Some fundamental findings of statistical seismology have also been
reported in compressive-failure experiments. First, the
Gutenberg-Richter law \cite{Utsu_GR} states that the number of
earthquakes as a function of their radiated energy $E$ decreases as a
power law, i.e., $p(E)dE \propto E^{-\epsilon} dE$ (with
$\epsilon=1+2b/3$ and $b$ close to 1). Numerous experiments on
compressive failure report power-law distributions in some measure of
the size of the events
\cite{Mogi62,Main,Alava_review,Davidsen_fracture}; however, there is
considerable scatter in the power-law exponents, which in addition can
either decrease with the evolution of the damage \cite{Alava_review}, or
show not so simple variations \cite{Main}. In general, there is a strong
influence of the external variables of the experiment, {mainly on
applied stress} \cite{Main}. Nevertheless, it is possible that some of
the early results are artifacts due to low counts and poor statistical
analysis.

The existence of power-law distributions and therefore of scale
invariance has led some authors to relate fracture with a second-order
phase transition \cite{Alava_review,Bonamy,Girard}, although others
point towards a first-order transition \cite{Alava_review,Rundle_review}, a debate
that replicates in earthquakes
\cite{Bak_book,Sornette_critical_book,Rundle_review,Ben_Zion_review}. In
any case, the broad range of responses triggered by the usual slow
perturbation is the signature of crackling noise \cite{Sethna_nature} (a
characterization that does not depend on the underlying mechanisms
generating the {output} of the system).

The (modified) Omori's law \cite{Utsu_omori} accounts for the fact
that the number of earthquakes per unit time decreases as a power law
since the sudden rise of activity provoked by a ``mainshock'', with an
exponent $p$ around 1. The counterparts of this law in fracture have
some problems of interpretation (whole rupture of the sample is the
mainshock \cite{Scholz68} versus similarity should hold also for
microfracturing bursts \cite{Hirata87}). Further, sometimes it is not
possible to distinguish the decay from an exponential form
\cite{Mogi62,Hirata87}, or the resulting $p$ is far from 1, although it
has been claimed that the $p-$exponent decreases as the experiment
progresses \cite{Hirata87}.

Time between consecutive events, or waiting times, have also been
measured in experiments under compression \cite{Mogi62,Alava_review}.
The Omori's law implies that the probability density of these times
should also follow a power-law decay with an exponent close to 1
\cite{Corral_Christensen}. However, the reciprocal is not true, since
power-law waiting times do not necessarily imply an underlying Omori's
law and therefore they are not a proof of the fulfillment of this law.

A coherent picture of waiting times in statistical seismology did not
start to consolidate until Bak et al. proposed their unified scaling law
\cite{Bak.2002}, measuring waiting times above a minimum energy in
different regions together. All the dependence on the size of the
regions and on the minimum energy turned out to be governed solely by a
unique parameter: the mean seismic activity rate $\langle r \rangle $,
in such a way that the waiting time probability density fulfills a
scaling law, $D(\delta) = \langle r \rangle \Phi(\langle r \rangle
\delta)$, with $\delta$ the waiting time and the scaling function $\Phi$
showing a power-law decay with exponent $1-\nu$ around 1 for small
arguments and another power law with exponent $2+\xi$ above 2 for large
arguments \cite{Corral_physA.2004}. Although the first exponent is a
consequence of the Omori's law, the second one is genuinely new, related
with the distribution of background seismic rates
\cite{Corral_Christensen}.

Compression experiments have shown good agreement with a restricted
version of this law \cite{Davidsen_fracture}, which considers the
special case of a single spatial region and a regime of stationary
seismicity (eliminating time periods with Omori-like decay
\cite{Corral_prl.2004}). In this case the scaling function turns out to
be well approximated by a flatter power-law decay (with $1-\nu$ around
0.3), followed by an exponential decay
\cite{Corral_prl.2004,Saichev_Sornette_times}.

Finally, another fundamental statistical law of seismic occurrence is
the productivity law \cite{Helmstetter03}, which establishes that the
rate of earthquakes (i.e., aftershocks) triggered by a mainshock of
energy $E$ is proportional to $E^{2 \alpha/3}$, with $\alpha \simeq
0.8$. As far as we know this law has not been reproduced in brittle
fracture experiments but in plastic deformation \cite{Weiss_Miguel}.

Therefore, there is no single compressive-failure experiment that
reproduces simultaneously the above mentioned fundamental laws of
statistical seismicity (Gutenberg-Richter, Omori, productivity, and the
unified waiting-time scaling law). The situation for tensile failure and
other types of tests is analogous
\cite{Bonamy,Alava_review,Mogi67,Astrom}, although the results of Ref.
\cite{Grob} are particularly notable, including spatial measurements.

In this Letter we report on the failure under compression of a highly
porous material, showing that the four main laws of statistical
seismicity hold, with unprecedented statistics, and with robust
exponents across different experiments. In contrast to the other laws,
the unified scaling law, which yields the best quantitative agreement
with earthquakes, is not stationary but arises from the temporal
variations of the activity rate.

We perform uniaxial compression experiments of Vycor, a mesoporous
silica ceramics (40\% porosity), loaded at a constant compression rate
$R$ {for three different experiments at} $R = $ 0.2, 1.6, and 12.2 kPa/s
(considering that the section of the sample keeps constant). Compression
is applied without lateral confinement until the shrinkage of the
samples is above 20\%, leading to multifragmentation. statistics.
Simultaneous recording of Acoustic Emission (AE) is performed by using a
detector coupled to the upper compression plate. The signal is
preamplified (60 dB), band filtered (between 20 kHz and 2 MHz) and
analyzed by means of a PCI-2 acquisition system from
EurophysicalAcoustics (Mistras Group) working at 1 MSPS. An AE avalanche
event starts at the time $t_i$ when the {preamplified} signal $V(t)$
crosses a fixed threshold of 26 dB, and finish when the signal remains
below threshold for more than 200 $\mu$s. The energy $E_i$ {associated
to} each event $i$ is computed as the integral of $V^2(t)$ for the
duration of the event divided by a reference resistance. More details of
the experiment can be found in Ref.~\cite{Salje}.

\begin{figure}[htb] 
\begin{center}
\epsfig{file=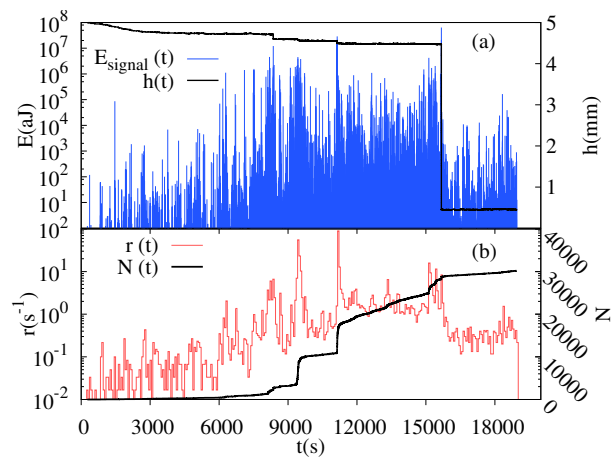,width=6cm,angle=270,clip=}
 \end{center} 
\caption{\label{FIG1}
(color online) (a) Example of {the outcome of} a compression experiment
{at} $R=1.6$ kPa/s, showing the change in the specimen's height $h$
versus time (proportional to stress) and the energy of the AE
avalanches, in logarithmic scale. (b) Time evolution of the AE activity
rate and of the total number of events.
}
\end{figure}

Fig.~\ref{FIG1}(a) shows an example of the raw results for the experiment at
$R=1.6$ kPa/s. The jerky evolution of the specimen's height is apparent,
as well as the broad range of values of the event energy detected at the
transducer. Another view of this intermittent dynamics is provided in
Fig.~\ref{FIG1}(b) by the AE activity rate $r(t)$ (counting events every 60 s)
and the cumulative number of events, $N(t)=\int_0^t r(t) dt$. Despite an
apparent correlation between the most energetic events and large changes
in height, one observes also regions with high acoustic activity not
associated with noticeable sample shrinkage.
%

\begin{figure}[htb]
\begin{center}
\epsfig{file=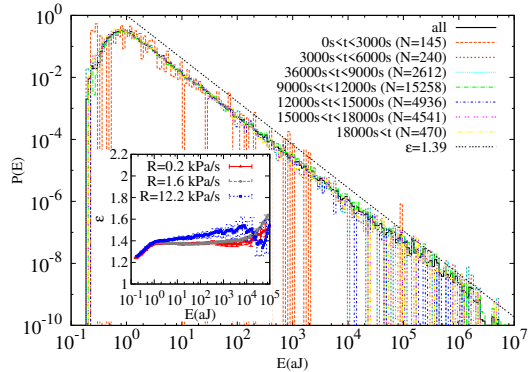,width=5cm,angle=270,clip=}
\end{center} 
\caption{\label{FIG2} 
(color online) Distribution of avalanche energies during the full
experiment with $R=1.6$ kPa/s and during 7 different subperiods. The
line shows the behavior corresponding to $\epsilon=1.39$. The inset
shows the ML-fitted exponent $\epsilon$ as a function of a lower
threshold $E_{min}$ for the three experiments.
} 
\end{figure}

Fig.~\ref{FIG2} shows the histograms that estimate the probability
densities of the energies\cite{Salje,Baro_Vives}, considering time
windows of $3 \times 10^3$s. All the distributions show a power-law
behavior $p(E) \propto E^{-\epsilon}$, with an exponent in the range
$\epsilon=1.40 \pm 0.05$, stable for the whole experiment; this is the
signature of a remarkable stationarity in the energy dissipation, which
appears as independent of applied stress, in contrast to previous works
\cite{Alava_review} (therefore, the apparent non-stationarity of $E$ in
Fig.~\ref{FIG1} is due to a much larger number of events in the central part).
The value of the exponent (obtained by maximum likelihood (ML)
estimation \cite{Baro_Vives}) holds for {about} 7 decades and is robust
against the thresholding of the data (fitting only values of $E$ larger
than $E_{min}$) and quite independent of $R$, as shown in the inset of
Fig.~\ref{FIG2} \cite{Salje,Baro_Vives,smalldev}. Although the resulting
exponent turns out to be below the most accepted value for earthquakes,
$\epsilon \simeq 1.67$, Kagan has noticed that this value is inflated
due to systematic biases and one could instead expect $\epsilon$ close
to 1.5 (i.e., $b\simeq 3/4$) \cite{Kagan_tectono10}. Reciprocally,
systematic biases of the energy cannot be completely ruled out in AE
experiments \cite{Alava_review,Bonamy}.
%

\begin{figure} 
\begin{center}
\epsfig{file=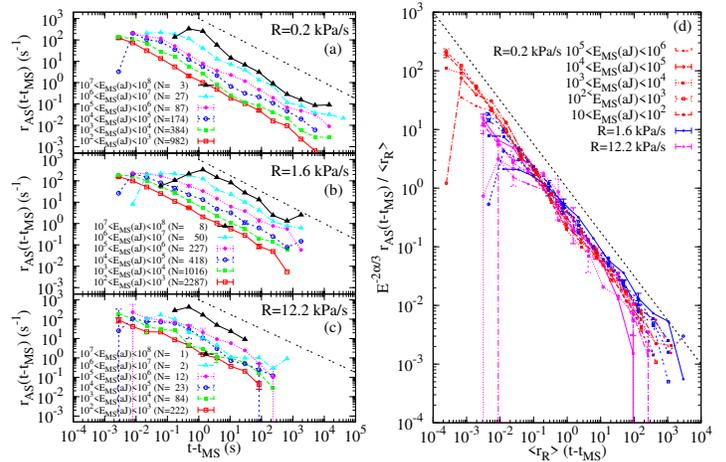,width=6.5cm,angle=270,clip=}
\end{center} 
\caption{\label{FIG3} 
(color online) Number of aftershocks per unit time, $r$, as a function
of the time distance to the main shock. MS are defined as the events in
the energy range indicated by the legend. $n$ values indicate the number
of sequences analyzed for each range. The dashed line indicates the
Omori's behavior with slope $-0.75$. Rescaled Omori plot showing the
fulfillment of the productivity law, with $\alpha \simeq 0.5$.
} 
 \end{figure}

The next step in our analysis has been the computation of the number of
aftershocks (AS) in order to compare with Omori's law for earthquakes.
We have considered as mainshocks (MS) all the events with energies in a
certain predefined energy interval. After eachMS we study the sequence
of {subsequent} events until an event with an energy larger than the
energy of the MS is found, {which} finishes the sequence of AS. Then we
divide the time line from the MS towards the future in intervals, for
which we count the number of AS {in each of them}. Averages of the
different sequences corresponding to all MS in the same energy range are
performed, normalizing each interval by the number of sequences that
reached such a time distance. The results presented in Figs.
~\ref{FIG3}(a-c) show that the tendency to follow Omori's law is clear,
in some cases for up to 6 decades, with an exponent $p=0.75\pm 0.10$.
(compare with Ref. \cite{Sornette_Ouillon}). Foreshocks, obtained in an
analogous way, show a similar behavior, with a slightly smaller value of
$p$.
%

The previous Omori's plot allows also to estimate the exponent $\alpha$
of the productivity law, by rescaling the vertical axis with
$E^{2\alpha/3}$, finding the optimum $\alpha$ which leads to the
collapse of the data; i.e., $r_{AS}/E^{2\alpha/3}$ should be only a
function of the time since the mainshock. The results in
Fig.~\ref{FIG3}(d) show that $\alpha=0.5 \pm 0.1$. This is again
somewhat smaller than the counterpart for earthquakes, but the drift is
compatible with the one found for the energy distribution, in other
words, the ratio of exponents $(\epsilon-1)/\alpha$ is the same.
Remarkably, a collapse can be obtained not only for mainshocks of
different energies in the same experiment but also across experiments
with different $R$, rescaling $r_{AS}$ as $r_{AS} E^{-2\alpha/3}/\langle
r_R \rangle$, and the time since the MS, $t-t_{MS}$, as $(t-t_{MS})
\langle r_R \rangle $, with $\langle r_R \rangle$ giving the mean number
of events per unit time (see the figure).

These results already suggest that there is a certain similarity in the
correlation between avalanches that extends from geophysical scales of
the order of hundreds of km to our small samples with cracks much
smaller than the mm scale. To deepen into the comparison we have
proceeded to the analysis of the interevent or waiting times, defined as
$\delta_{j}=t_{j} - t_{j-1}$, with $j$ labeling only the events with
energy larger than a given $E_{min}$. The estimations of the
waiting-time probability densities, $D(\delta;E_{min})$, for different
$E_{min}$ and different experiments are shown in Fig.~\ref{FIG4}(a),
displaying a power-law decay with exponent $1-\nu=0.93 \pm 0.05$ for most of
the time range, as implied by the Omori's law. In order to
compare the shape of the distributions we rescale the axes as $\langle r
(E_{min})\rangle \delta $ and $D(\delta;E_{min},R)/\langle r
(E_{min})\rangle$, with $\langle r (E_{min})\rangle $ giving the mean
number of events per unit time with $E \ge E_{min}$. Fig.~\ref{FIG4}(b)
shows {how the different distributions collapse into a single one,
signaling the existence of a scaling law; for a single experiment, as
the activity rate verifies the Gutenberg-Richter law, the collapse
``unifies'' this law with the temporal properties \cite{Bak.2002}. For
different experiments the collapse implies the similarity versus the
compression rate $R$. Moreover, the plot also shows that a second power
law emerges for the {rightmost} tail of the distributions, with an
exponent $2+\xi=2.45 \pm 0.08$ \cite{footnote}.

To make clear the correspondence with earthquakes Fig.~\ref{FIG4}(b) also
includes seismic data for different spatial windows in Southern
California \cite{Bak.2002,Corral_physA.2004}. Although the previously
reported value of $\xi$ for earthquakes \cite{Corral_physA.2004} is a
bit smaller than for the experiment, the similarity is remarkable,
taking into account that the earthquake measurements are taken over
different spatial windows, whereas for the AE data we do not have access
to such degrees of freedom.
%
\begin{figure}[htb] 
\begin{center} 
\epsfig{file=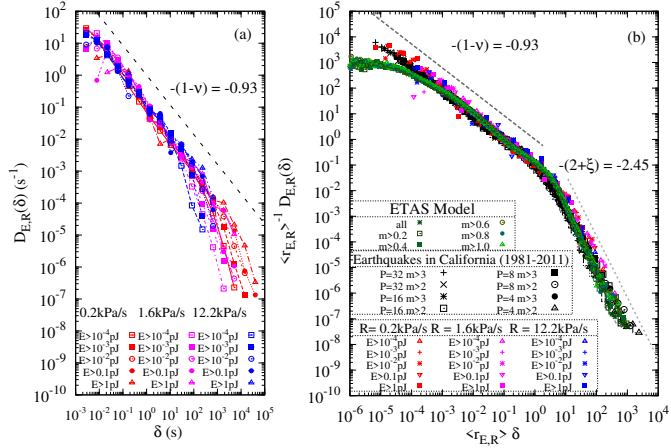, width=6cm ,angle=270,clip=} 
\end{center} 
\caption{\label{FIG4} 
(color online) (a) Distribution of waiting times for different values of
$E_{min}$ and the compression rate $R$. (b) The same data under
rescaling, including also the results of the ETAS model and earthquakes
from Southern-California divided into $P\times P$ regions
\cite{Bak.2002,Corral_physA.2004} for the period Jan 1984 -- Jun 2011.
}
\end{figure}

How can we get then essentially the same behavior in such different
situations? The answer lies in the variations of the activity rate. Let
us consider a single Omori sequence, for which the waiting-time density
depends on the background activity rate $\mu$ through a scaling form
\cite{Corral_Christensen},
\begin{equation}
D(\delta | \mu) =\frac \mu {( \mu \delta)^{1-\nu}} f(\mu \delta),
\end{equation}
where $\nu$ is close to 0 and $f$ can be a decreasing exponential, or
another function showing the same behavior at $0$ and $\infty$. If the
background rate is not fixed but evolves during the experiment, the
resulting density will be
\begin{equation}
D(\delta) 
\propto \int_{\mu_{min}}^{\mu_{max}} 
d\mu \rho(\mu) \mu D(\delta | \mu),  
\end{equation}
where $\rho(\mu)$ is the density of background rates. Substituting the
previous equation and considering that $\mu$ is distributed between
$\mu_{min}$ and $\mu_{max}$ with $\rho(\mu) \propto 1/\mu^{1-\xi}$ leads
to $D(\delta) \propto 1/\delta^{1-\nu} $ for $\delta \ll \mu_{max}^{-1}$
(because the rescaled integral goes to zero as $\delta^{1+\xi+\nu}$) but
$D(\delta) \propto 1/\delta^{2+\xi} $ for $\delta \gg \mu_{max}^{-1}$
(because the rescaled integral converges to a constant). This behavior
for $\rho(r)$ can arise from a time evolution of the form $\mu(t)
\propto t^{1/\xi}$, as $\rho(\mu) \propto |dt/d\mu(t)|$
\cite{Corral_Christensen}. So, when the background rate varies across
different scales (as in Fig.~\ref{FIG1}(b)) and this takes place through
a power law, a second power law arises in $D(\delta)$. The experimental
outcome suggests {then} $\xi \simeq 0.5$. We have simulated the Epidemic
Type Aftershock (ETAS) model \cite{Helmstetter_Sornette_jgr02},
defined by the fact that each earthquake $i$, with a Gutenberg-Richter
energy, triggers a sequence with a rate equal to $K
{E_i}^{2\alpha/3}/(c+t-t_i)^{1+\theta}$, and the overall rate is the
linear superposition of these rates plus a background rate. The
``microscopic'' exponent $1+\theta$ corresponds to an observable
$p=1-\theta$ \cite{Helmstetter_Sornette_jgr02}. Using as input the
experimental values of $\epsilon$, $p$, and $\alpha$, together with
$c=0.001$ s, and $\mu$ increasing slowly as $\mu(t) \propto 1-\cos
\omega t$ (essentially a power law with $\xi=1/2$) we obtain very good
concordance with the previous calculations (see Fig. \ref{FIG4}(b)) when
the branching ratio (given by $Kb/(\theta c^\theta (b-\alpha))$) is very 
close to criticality, i.e.
0.99. Also, the measurement of $r(t)$, using different time intervals,
leads to a distribution with a power-law tail of the form $1/\sqrt{r}$
for small $r$ (not shown). This explanation could hold also for Ref.
\cite{Santucci_Vanel_Ciliberto}.

In summary, we have presented experimental results on the compression of
a highly porous material, obtaining good fulfillment of some fundamental
laws of statistical seismology. Laws involving the measurement of energy
and the Omori's law show some bias in the exponent with respect the
earthquake case, whereas for the unified scaling law the quantitative
agreement is much better. As our experiment does not allow the
measurement of the location of the events, it has been not possible to
test laws regarding spatial properties, which constitute also an
important body of knowledge for the characterization of seismicity
\cite{Grob}. However, the validity of the unified scaling law in our
experiments is associated to temporal variations of the background
activity rate, rather than to spatial variations.

We acknowledge financial support from the Spanish Ministry of Science
(MAT2010-15114), and the Austrian Science Fund (FWF) P23982-N20.


\begin{thebibliography}{99}
\bibitem{smalldev}
The small deviation corresponding to the largest compression rate
$R=12.2$ kPa/s can be due
to {some non-trivial effects of overlapping} and/or distortion of the
avalanches.
\end{thebibliography}
\end{document}